\documentclass[12pt]{article}
\usepackage{graphicx}
\usepackage{graphics}
\usepackage{url}
\usepackage{cancel}
\usepackage{amsmath}

\newcommand\pubnumber{CERN-PH-TH-2014-166}
\newcommand\pubdate{\today}

\def\napoli{CERN Theory Division,\\ CH-1211 Geneva 23,\\ SWITZERLAND}
\def\support{\footnote{Work supported by a Marie Curie Intra-European Fellowship of the European Community's 7th Framework Programme under contract number (PIEF-GA-2012-326948).}}

\def\Title#1{\begin{center} {\Large #1 } \end{center}}
\def\Author#1{\begin{center}{ \sc #1} \end{center}}
\def\Address#1{\begin{center}{ \it #1} \end{center}}

\newcommand\pubblock{\rightline{\begin{tabular}{l} \pubnumber\\
         \pubdate  \end{tabular}}}
\newenvironment{Abstract}{\begin{quotation}  }{\end{quotation}}
\newenvironment{Presented}{\begin{quotation} \begin{center} 
             PRESENTED AT\end{center}\bigskip 
      \begin{center}\begin{large}}{\end{large}\end{center} \end{quotation}}
\def\Acknowledgements{\bigskip  \bigskip \begin{center} \begin{large}
             \bf ACKNOWLEDGEMENTS \end{large}\end{center}}

\def\beq{\begin{equation}}
\def\eeq#1{\label{#1}\end{equation}}
\def\eeqn{\end{equation}}

\def\beqa{\begin{eqnarray}}
\def\eeqa#1{\label{#1}\end{eqnarray}}
\def\eeqan{\end{eqnarray}}

\let\bar=\overbar

\def\bra#1{\left\langle{ #1} \right|}
\def\ket#1{\left| {#1} \right\rangle}

\def\Dslash{\not{\hbox{\kern-4pt $D$}}}
\def\dslash{\not{\hbox{\kern-2pt $\del$}}}

\def\msb{{\bar{\ssstyle M \kern -1pt S}}}

\begin{document}
\begin{titlepage}
\pubblock

\vfill
\Title{Challenges for New Physics in the Flavour Sector}
\vfill
\Author{Andreas Crivellin\support}
\Address{\napoli}
\vfill
\begin{Abstract}
In these proceedings I present a personal perspective of the challenges for new physics (NP) searches in the flavour sector. Since the CKM mechanism of flavour violation has been established to a very high precision, we know that physics beyond the Standard Model can only contribute sub-dominantly. Therefore, any realistic model of physics beyond the Standard Model (SM) must respect the stringent constrains from flavour observables like $b\to s \gamma$, $B_s\to\mu^+\mu^-$, $\Delta F=2$ processes etc., in a first step. In a second step, it is interesting to ask the question if some deviations from the SM predictions (like the anomalous magnetic moment of the muon or recently observed discrepancies in tauonic $B$ decays or $B\to K^*\mu^+\mu^-$) can be explained by a model of NP without violating bounds from other observables.
\end{Abstract}
\vfill
\begin{Presented}
Flavor Physics and CP Violation (FPCP-2014),\\ Marseille, France, \\May 26-30 2014\end{Presented}
\vfill
\end{titlepage}
\def\thefootnote{\fnsymbol{footnote}}
\setcounter{footnote}{0}

\section{Introduction}
\begin{figure}[htb]
\centering
\includegraphics[height=4in]{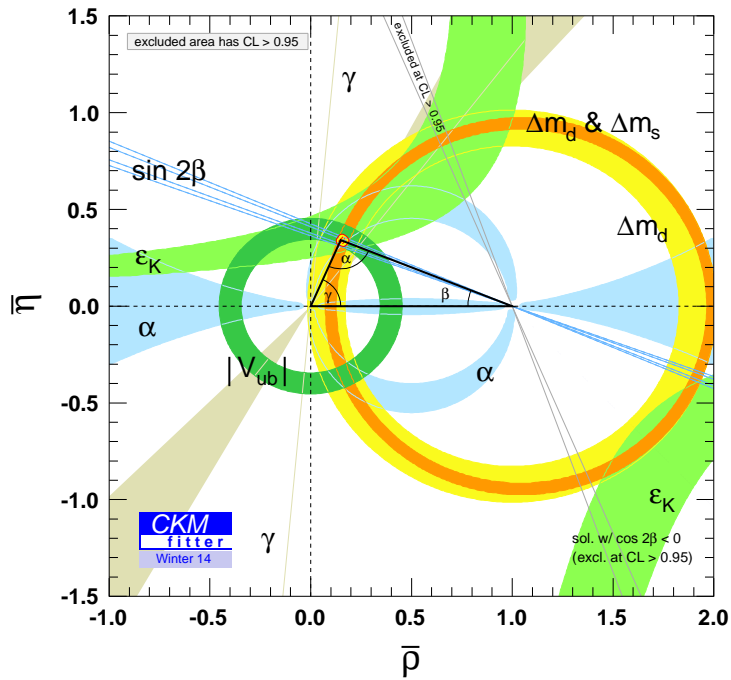}
\caption{Global fit to the CKM matrix performed by the CKMfitter collaboration \cite{CKMfitter} shown in the $\bar \rho$--$\bar \eta$ plane defined as $\bar \rho +i \bar\eta=-\dfrac{V_{ub}^*V_{ud}}{V_{cb}^*V_{cd}}$. One can see that the allowed regions from the different observables overlap in a small region, leaving limited space for NP contributions. For a similar analysis of the UTfit collaboration see \cite{UTfit}.}
\label{fig:CKMfit}
\end{figure}

The CKM mechanism of flavour violation has been established by the $B$ factories BELLE and BABAR and received further confirmation by the LHCb experiment. Global fits to the CKM matrix \cite{Charles:2004jd,Bona:2005vz} show that there is in general a very good agreement between the different observables and that new physics (NP) contributions can only be of the order of $10\%$ (see Fig.~\ref{fig:CKMfit}): The global CKM fit includes the tree-level determinations of the CKM elements $V_{us}$, $V_{cb}$ and $V_{ub}$ as well as information from $K-\overline{K}$, $B_s-\overline{B}_s$ and $B_d-\overline{B}_d$ mixing. In addition, also the measurements of processes not included in the CKM fit like $B_{s,d}\to \mu^+\mu^-$ and $b\to s,d\gamma$ agree very well with the Standard Model (SM) predictions.

Therefore, the challenges for NP in the flavour sector are the following: In a first step, any model of NP must respect the stringent constraints from flavour observables. In a second step one can examine if the model under consideration can explain deviations from the SM without violating bounds from other observables. It is interesting to note that in general it is rather difficult to construct a model which satisfies the stringent constraints from FCNC processes while still being capable of explaining some deviations. While step one can be fulfilled by assuming some symmetry or alignment with the SM, suppressing flavour effects, step 2 often requires a rather generic flavour structure leading in many cases to tensions. 

Among the many flavour observables which are in very good agreement with the SM predictions we will consider:
\begin{itemize}
	\item $\Delta F=2$ processes: $B_s-\overline{B}_s$, $B_d-\overline{B}_d$, $K-\overline{K}$ and $D-\overline{D}$ mixing.
	\item Radiative $B$ meson decays: $b\to s\gamma$ and $b\to d\gamma$.
	\item Neutral meson decays to muon pairs: $B_s\to\mu^+\mu^-$, $B_d\to\mu^+\mu^-$, $K_L\to\mu^+\mu^-$ and $D\to\mu^+\mu^-$.
	\item Lepton flavour violating observables: $\ell_i\to \ell_f \gamma$, $\mu\to e$ conversion and $\ell_i\to\ell_f\ell_j\ell_j$.
\end{itemize}
The stringent bounds from these observables must be respected by any reasonable NP model. Many possibilities how to ensure that NP models satisfy these bounds have been proposed, among them minimal flavour violation is probably the one most often used in the literature \cite{MFV}.

On the other hand, there are some observables in which some tensions with the SM have been observed, among them:
\begin{itemize}
	\item Tauonic $B$ decays: $B\to\tau\nu$, $B\to D\tau\nu$ and $B\to D^*\tau\nu$.
	\item $B\to K^*\mu^+\mu^-$ and ${\rm Br}[B\to K\mu^+\mu^-]/{\rm Br}[B\to Ke^+e^-]$.
\end{itemize}
Here we want to address the question which model of NP is capable to explain theses deviations from the SM. 

There are also tensions among the different determinations of $V_{ub}$ and $V_{cb}$ from different inclusive and exclusive processes (see for example \cite{Beringer:1900zz,Ricciardi:2013cda} for a review). However, it is essentially ruled out that the current data for $V_{cb}$ can be explained by NP contributions and also in the case of $V_{ub}$\footnote{For $V_{ub}$ it is still possible that a right-handed $W$ couplings (as proposed in \cite{Chen:2008se}) could explain the differences between the inclusive and exclusive determinations\cite{Bernlochner:2014ova}} it is rather likely that the differences are due to underestimated errors in the theory predictions \cite{Crivellin:2014zpa}. Therefore, we do not consider these discrepancies in the following.

\subsection{The 2HDM and the MSSM in the decoupling limit}

In the next section we will discuss several decays putting stringent constraints on physics beyond the SM and Sec.~\ref{sec:deviations} discusses some observables in which deviations for the SM predictions have been observed. We will illustrate the impact of these observables on some selected on NP models. As a specific example we consider a 2HDM \cite{Lee:1973iz} (for a review see for example \cite{Branco:2011iw}) with generic Yukawa structure (which is the decoupling limit of the MSSM). Here we introduce a second Higgs doublet and obtain four additional physical Higgs particles (in the case of a CP conserving Higgs potential): the neutral CP-even Higgs $H^0$, a neutral CP-odd Higgs $A^0$ and the two charged Higgses $H^{\pm}$. In addition, if we allow for a generic flavour structure we have the non-holomorphic couplings which couple up (down) quarks to the down (up) type Higgs doublet. Following the notation of Ref.~\cite{Crivellin:2013wna}
\begin{eqnarray}
\renewcommand{\arraystretch}{2.2}
\begin{array}{l}
\mathcal{L}^{eff}  = \bar u_{f\;L}^{} V_{fj} \left( {\dfrac{{m_{d_i} }}{{v_d }}\delta_{ij}H_d^{2\star}  - \epsilon_{ji}^{d} \left( {H_u^1  + \tan \left( \beta  \right)H_d^{2\star} } \right)} \right) d_{i\;R}  \\ 
\phantom{\mathcal{L}^{eff}  =}  
+ \bar d_{f\;L} V_{j f}^{\star} \left( {\dfrac{{m_{u_j} }}{{v_u }}\delta_{ij}H_u^{1\star}  - \epsilon_{ji}^{u} \left( {H_d^2  + \cot \left( \beta  \right)H_u^{1\star} } \right)} \right) u_{i\;R}  \\ 
\phantom{\mathcal{L}^{eff}  =}  
- \bar d_{f\;L}  \left( {\dfrac{{m_{d_i} }}{{v_d }}\delta_{fi}H_d^{1\star}  + \epsilon_{fi}^{d} \left( {H_u^2  - \tan \left( \beta  \right)H_d^{1\star} } \right)}  \right) d_{i\;R}  \\ 
\phantom{\mathcal{L}^{eff}  =}  - \bar u_{f\;L}^a \left( {\dfrac{{m_{u_i} }}{{v_u }}\delta_{fi}H_u^{2\star}  + \epsilon_{fi}^{u} \left( {H_d^1  - \cot \left( \beta  \right)H_u^{2\star} } \right)} \right)u_{i\;R} \,+\,h.c.  \\ 
 \end{array}
\label{L-Y-FCNC}
\end{eqnarray}
Here $\epsilon^q_{ij}$ parametrizes completely flavour-chaining neutral currents. In the MSSM at tree-level $\epsilon^q_{ij}=0$ (which corresponds to the 2HDM of type II) and flavour changing neutral Higgs couplings are absent. However, these couplings are generated at the loop level\cite{Banks:1987iu}. The resulting expressions are non-decoupling and depend only on the ratios of SUSY parameters (for a complete one-loop analysis see \cite{Crivellin:2010er} and for the 2-loop SQCD corrections see \cite{Crivellin:2012zz}\footnote{In the flavour-conserving case the 2-loop corrections were calculated in Ref.~\cite{Noth:2010jy}}). Since the dependence is only on the ratio of SUSY masses, these effects are non-decoupling and even allow for the possibility that the light fermion masses arise entirely from $\epsilon^f_{ij}$ \cite{Lahanas:1982et}.

\section{Selected flavour-processes and their constraints on new physics}

In this section we review some selected flavour observables and highlight their impact on models of NP, i.e. how they constrain physics beyond the SM. 

\boldmath
\subsection{$\Delta F=2$ processes}
\unboldmath
	
$\Delta F=2$ processes are still one of the most constraining processes for NP (see for example \cite{Bona:2007vi} for a model-independent analysis and \cite{Lenz:2010gu} for an overview on $B_q-\bar B_q$ and $K-\overline{K}$ mixing) since they scale like $\delta^2/\Lambda^2$ while the other flavour observables scale like $\delta/\Lambda^2$. Here $\delta$ stands for a generic flavour violating parameter and $\Lambda$ is the scale of NP. Especially the constraints from $K-\overline{K}$ and $D-\overline{D}$ mixing are very stringent and Kaon mixing puts extremely stringent constrainrs on CP violating NP.

\begin{figure}[htb]
\centering
\includegraphics[height=2.5in]{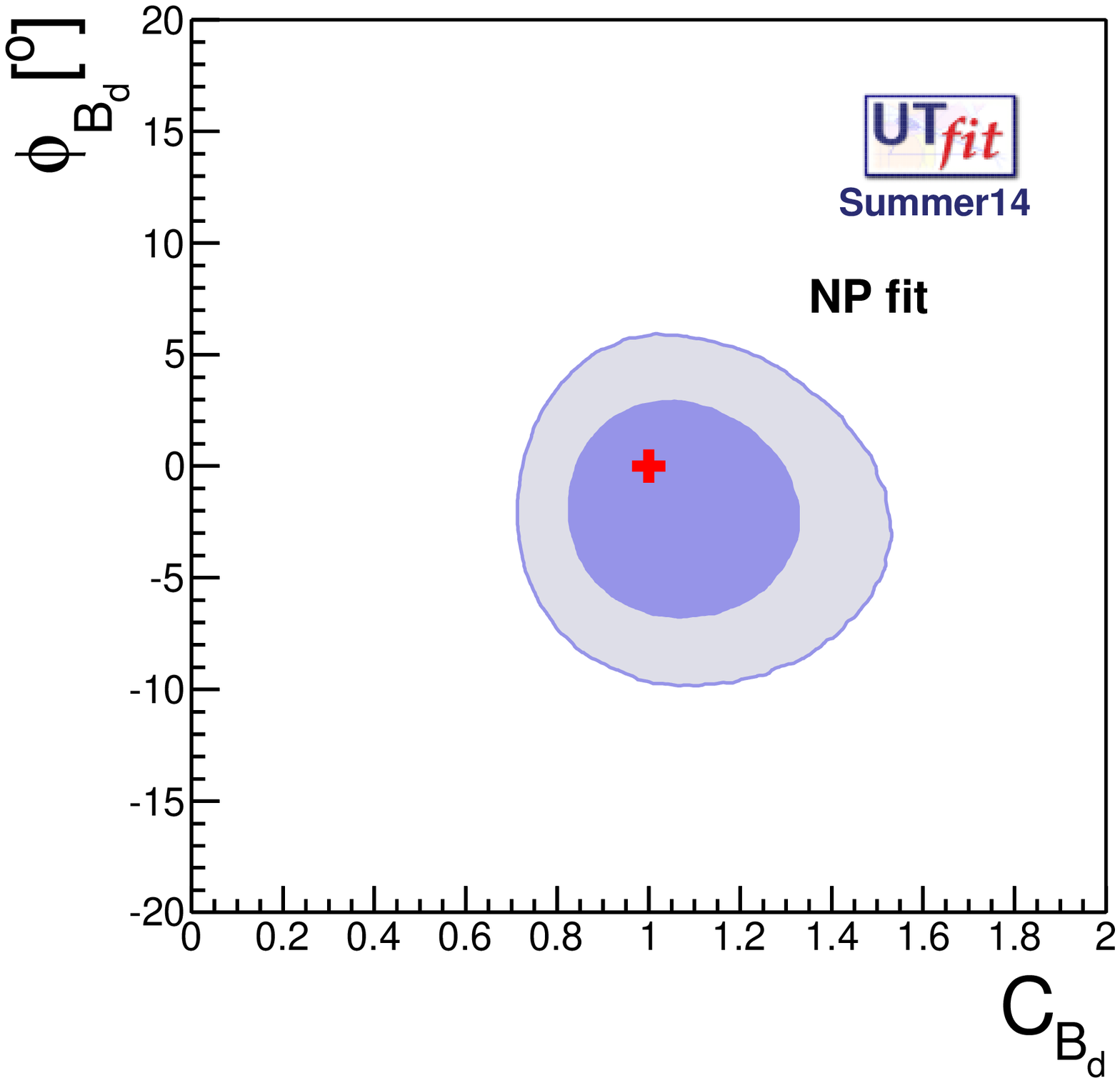}
\includegraphics[height=2.5in]{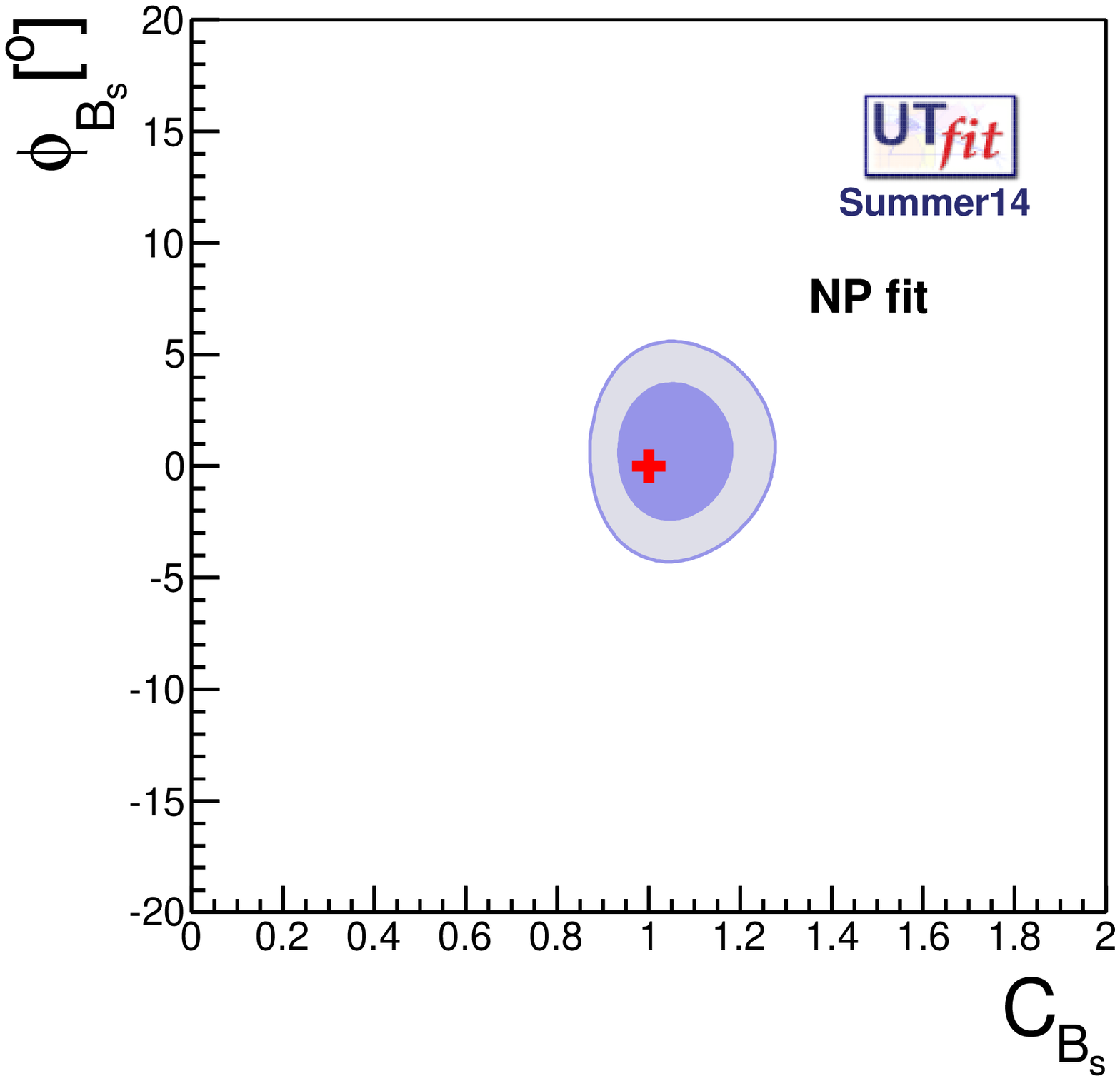}
\caption{Analysis of NP in $B_q-\overline{B}_q$ mixing of the UTfit collaboration \cite{UTfit}. For a similar analysis of the CKMfitter collaboration see \cite{CKMfitter}. The SM point (red cross) corresponds to $C_{B_q}=1$ and $\phi _{B_q }=0$. }
\label{NPinMixing}
\end{figure}

The current situation concerning the experimental and theoretical values is the following: For $K-\overline{K}$ mixing the SM prediction was calculated at NLO in Ref.~\cite{Buras:1990fn,Herrlich:1993yv} and (for the relevant charm contributions) at NNLO in Ref.~\cite{Brod:2010mj}. 
\begin{equation}
|\epsilon_K|_{\rm SM} = 1.81(28) \times 10^{−3}\,,\qquad \Delta M^{\rm SD\; SM}_K = 3.1(1.2) \times 10^{−15} \,{\rm GeV}\,.
\end{equation}
Here $\Delta M^{\rm SD\; SM}_K$ only contains the calculable short distance SM contribution. This has to be compared to the experimental value \cite{Beringer:1900zz}
\begin{equation}
|\epsilon_K|_{\rm exp} = (2.228 \pm 0.011)\times 10^{−3}\,,\qquad\Delta M^{\rm exp}_K=3.48\pm0.06\times 10^{-15}\, {\rm GeV}\,.
\end{equation}
Concerning the SM predictions for $B_s-\overline{B}_s$ and $B_d-\overline{B}_d$ mixing (calculated in Ref.~\cite{Buras:1990fn}) the latest numerical update \cite{Lenz:2006hd} gives
\begin{eqnarray}
\Delta M_d^{\rm SM} &=& 0.502 \,{\rm ps}^{-1}\,,\qquad \phi_d^{\rm SM} = (-−10.1^{+3.7}_{-−6.3}) \times 10^{-−2}\,.\\
\Delta M_s^{\rm SM} &=& 17.24 \,{\rm ps}^{-1}\,,\qquad \phi_s^{\rm SM} = (7.4^{+0.8}_{-−3.2}) \times 10^{-−3}\,.
\end{eqnarray}
This is in very good agreement with the experimental values from LHCb \cite{Aaij:2013mpa} and CDF \cite{Abulencia:2006ze} as well with the HFAG average for $B_d-\overline{B}_d$ mixing. In general, one can parametrize the NP contribution to $B_q-\overline{B}_q$ mixing as
\begin{equation}
C_{B_q} e^{2 i \phi _{B_q } }  = |\Delta_q| e^{i \phi _{q }^{\Delta} }
  = \frac{\bra{B_q} H_{\rm eff} \ket{\bar B_q}}{
          \bra{B_q} H_{\rm eff}^{\rm SM} \ket{\bar B_q}}\,,
 \label{NP-B-mixing}
\end{equation}
where the first notation is used by the UTfit collaboration and the second one by CKMfitter.
The experimental information for CP violating quantities and the mass differences are correlated and one obtains the allowed regions in the plane shown in Fig.~\ref{NPinMixing}.

For $D-\overline{D}$ mixing the SM calculation for the mass difference is not reliable but it is sensible to assume that NP does not generate more that the total observed value. In addition the CP-violating phase in the SM is very small \cite{Bobrowski:2010xg} resulting in stringent bounds on NP derived from recent experimental data \cite{Bevan:2014tha}.

\begin{figure}[t]
\centering
\includegraphics[width=0.3\textwidth]{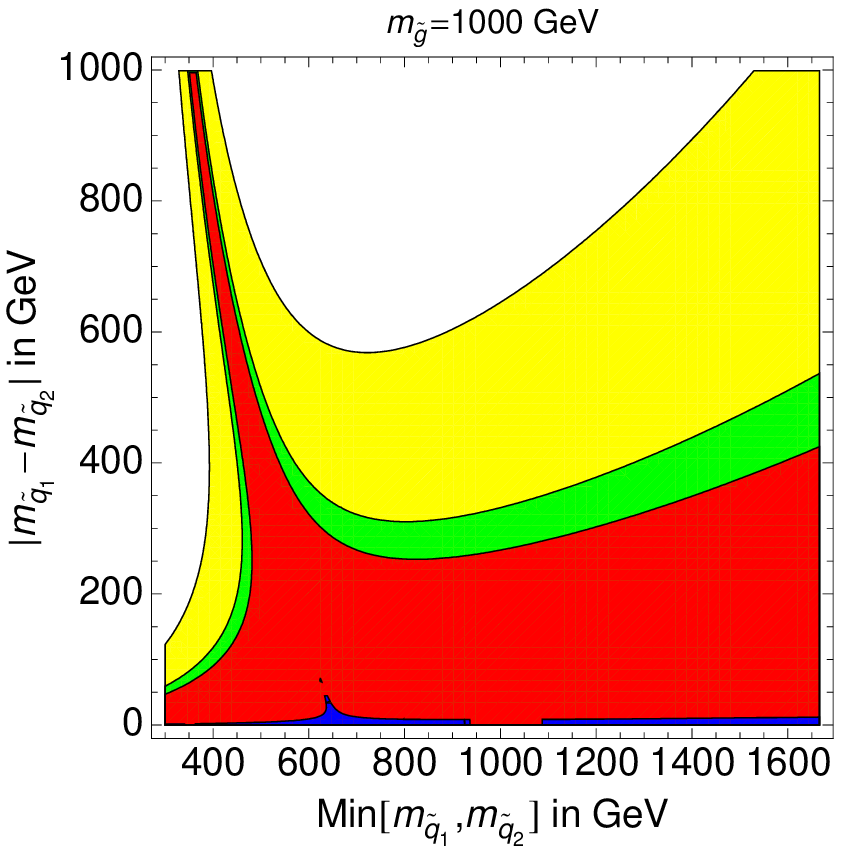}~~~~~
\includegraphics[width=0.3\textwidth]{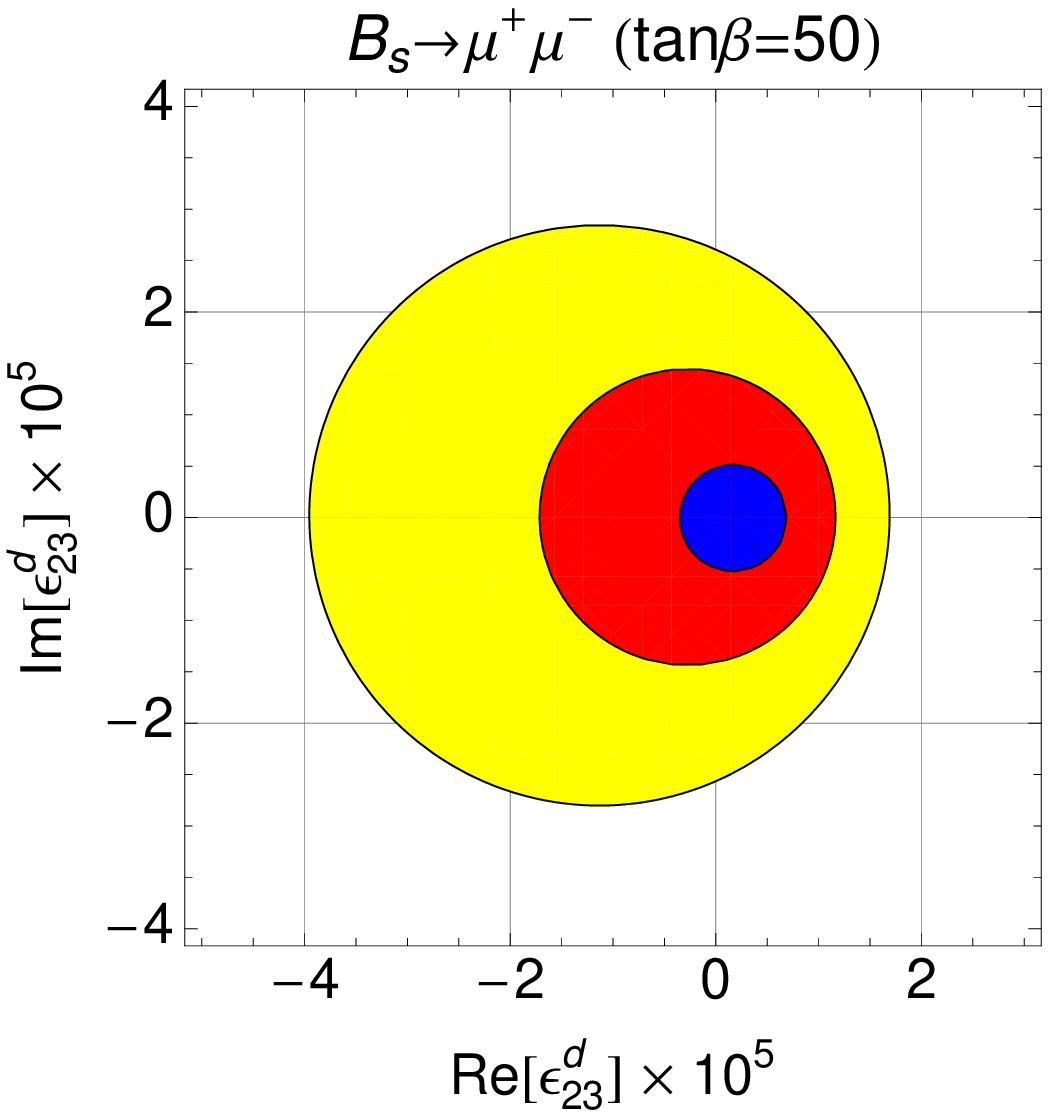}
\includegraphics[width=0.334\textwidth]{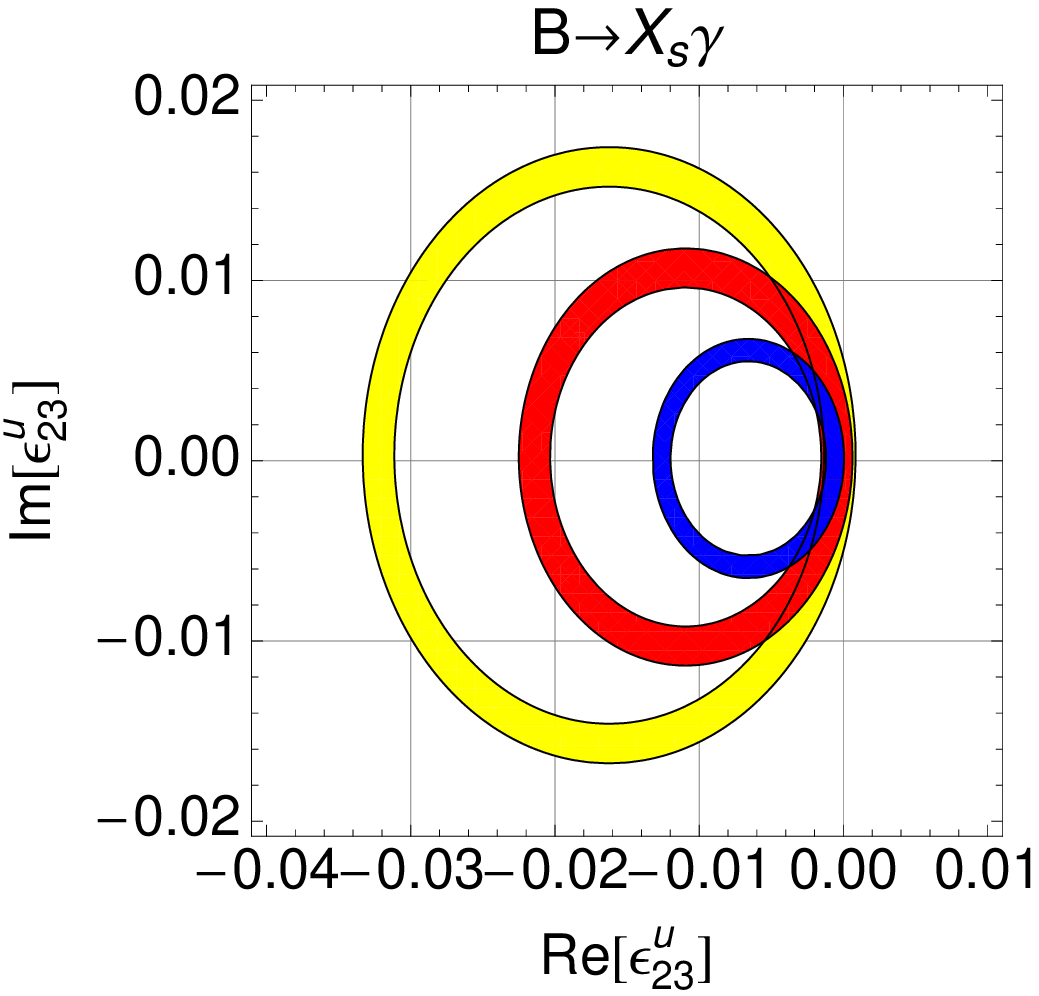}
\caption{Left: Allowed mass splitting between the first two generations of left-handed squarks for different gluino masses for $M_2=(\alpha_2/ \alpha_s) m_{\tilde g}\cong 0.35$ taken from Ref~\cite{Crivellin:2010ys}. Yellow (lightest) corresponds to the maximally allowed mass splitting assuming an intermediate alignment of $m^2_{\tilde{q}}$ with $Y_u^{\dagger}Y_u$ and $Y_d^{\dagger}Y_d$. The green (red) region is the allowed range assuming an diagonal up (down) squark mass matrix. The blue (darkest) area is the minimal region allowed in which the off-diagonal element carries a maximal phase. Middle: Allowed regions in the complex $\epsilon^{d}_{23}$--plane from $B_s\to\mu^+\mu^-$ for $\tan\beta=50$ and $m_{H}=700\mathrm{~GeV}$ (yellow), $m_{H}=500\mathrm{~GeV}$ (red) and $m_{H}=300\mathrm{~GeV}$ (blue). Note that the allowed regions for $\epsilon^{d}_{32}$--plane are not full circles because in this case a suppression of ${\cal B}\left[B_{s}\to\mu^+\mu^-\right]$ below the experimental lower bound is possible. Right: Allowed regions for $\epsilon^{u}_{23}$ from $ B \to X_{s} \gamma$, obtained by adding the $2\,\sigma$ experimental error and theoretical uncertainty linear for $\tan\beta=50$ and $m_{H}=700 \, \mathrm{ GeV}$ (yellow), $m_{H}=500\, \mathrm{ GeV}$ (red) and  $m_{H}=300 \,\mathrm{ GeV}$ (blue). The middle and the right plot are taken from Ref.~\cite{Crivellin:2013wna}.}
\label{fig:Bstomumu}
\end{figure}

Therefore, $\Delta F=2$ processes put stringent constraints on physics beyond the SM. Especially combining $D-\overline{D}$ and $K-\overline{K}$ mixing is very powerful due to $SU(2)_L$ relations for NP \cite{Blum:2009sk}. As an example of how $\Delta F=2$ processes constrain models of NP we consider the MSSM. Here the mass splitting between the first two generation of left-handed squarks is limited by $K-\overline{K}$ and $D-\overline{D}$ mixing \cite{Nir:1993mx}. However, due to cancellations among the different contributions (gluino, chargino and neutralino) it has been shown that the mass splitting can still be sizable \cite{Crivellin:2010ys} (right plot of Fig.~\ref{fig:Bstomumu}). A mass splitting among the first two generations has interesting consequences for LHC searches \cite{Mahbubani:2012qq}.

\subsection{$B_q\to\mu^+\mu^-$}

Thanks to LHCb and CMS \cite{Aaij:2012nna} we know the branching ratio for $B_q\to\mu^+\mu^-$ now rather precisely and also the SM prediction has been improved recently \cite{Bobeth:2013uxa}:
\begin{align}
	{\rm Br}{\left[ {{B_s} \to {\mu ^ + }{\mu ^ - }} \right]_{\exp }} = \left( {2.9 \pm 0.7} \right) \times {10^{ - 9}}\,,\qquad {\rm Br}{\left[ {{B_s} \to \mu \mu } \right]_{SM}} = \left( {3.65 \pm 0.23} \right) \times {10^{ - 9}}\,.
\end{align}
\begin{align}
		{\rm Br}{\left[ {{B_d} \to {\mu ^ + }{\mu ^ - }} \right]_{\exp }} < 7.4 \times {10^{ - 9}}\,,\qquad {\rm Br}{\left[ {{B_d} \to \mu \mu } \right]_{SM}} = \left( {3.6^{+1.6}_{-1.4}} \right) \times {10^{ - 10}}\,,\\
\end{align}
Due to the good agreement with the SM we can place stringent bounds on models of NP, especially if the NP model possesses scalar currents \cite{Babu:1999hn}. 

As an example we consider again the 2HDM of type III. In the middle plot of Fig.~\ref{fig:Bstomumu} we show the constraints on the parameter $\epsilon^d_{23,32}$ which generate $B_s\to\mu^+\mu^-$ via a tree-level Higgs exchange. 
While the experimental bounds on $B_d\to\mu^+\mu^-$ are still weaker compared to the SM prediction but LHCb will further improve the experimental limit in the future. Also here stringent limits on $\epsilon^d_{13,31}$ can be obtained. Concerning $K_L\to\mu^+\mu^-$ the SM predictions is limited by hadronic uncertainties related to photon rescattering which complicates the extraction on bounds on NP \cite{Isidori:2003ts}. For  $D\to\mu^+\mu^-$ the SM predictions is not reliable but it is again sensible to assume that not more than the entire decay rate is generated by NP. Therefore, we can derive in an analogous way stringent constraints on $\epsilon^{u,d}_{12,21}$ as well. In summary, neutral meson decays to muons constrain all flavour-changing elements $\epsilon^d_{ij}$ and $\epsilon^u_{12,21}$ stringently.

\subsection{$b\to q\gamma$}

Concerning the radiative $B$ decays $b\to s\gamma$ and $b\to d\gamma$ the current experimental values \cite{Chen:2001fja} and theoretical predictions are given by:
\begin{eqnarray}
 {\rm Br}{\left[ {b \to s\gamma } \right]_{\exp }} =& \left( {3.43 \pm 0.21 \pm 0.07} \right) \times {10^{ - 4}}\,,\\
 {\rm Br}{\left[ {b \to s\gamma } \right]_{SM}} =& \left( {3.15 \pm 0.22} \right) \times {10^{ - 4}}\,, \\ 
 {\rm Br}{\left[ {b \to d\gamma } \right]_{\exp }} =& \left( {1.41 \pm 0.57} \right) \times {10^{ - 5}} \,,\\
 {\rm Br}{\left[ {b \to d\gamma } \right]_{SM}} =& 1.54_{ - 0.31}^{ + 0.26} \times {10^{ - 5}} \,.
\end{eqnarray}
Again, we observe a good agreement between theory predictions\footnote{The SM prediction for $b \to d\gamma$ is taken from \cite{Crivellin:2011ba} (based on the calculation of Ref.~\cite{Ali:1992qs}) while the value for $b\to s\gamma$ is taken from Ref.~\cite{Misiak:2006zs}. For the experimental value of $b\to s\gamma$ we used the average of \cite{Amhis:2012bh}.} and experiment. 

$b\to s\gamma$ has been used extensively to constrain models of NP, especially the MSSM and the 2HDM of type II \cite{Bertolini:1990if}. In fact, $b\to s\gamma$ still gives the best lower bound of $380$~GeV \cite{Hermann:2012fc} on the charged Higgs mass in the 2HDM of type II for low or moderate values of $\tan\beta$ (see Fig. \ref{fig:2HDMII}).

$b\to s\gamma$ can for example also be used to put bounds on $\epsilon^u_{23}$ originating from charged Higgs loop contributions. The results are shown in the right plot of Fig.~\ref{fig:Bstomumu}. Similar constrains apply for $\epsilon^u_{13}$ from $b\to d\gamma$.

\subsection{Lepton flavour violation}

Flavor-changing neutral current processes are strongly suppressed in the Standard Model (SM) and therefore sensitive even to small new physics (NP) contributions. Lepton flavor violation (LFV) is an especially promising probe of NP since in the SM with massive neutrinos all flavor-violating effects in the charged lepton sector are proportional to tiny neutrino masses.\footnote{For a review we refer to~\cite{Raidal:2008jk}.} For instance, the decay rates of heavy charged leptons into lighter ones are suppressed at least by $m_\nu^4/m_W^4$, where $m_\nu$ ($m_W$) is the neutrino ($W$-boson) mass. This leads to branching ratios of the order of $10^{-50}$, which are thus by far too small to be measurable in any foreseeable experiment. Therefore, any evidence of charged LFV would be a clear signal of physics beyond the SM. 

\begin{center}
\begin{table}[htdp]
\begin{minipage}{2in}
\centering \vspace{0.8cm}
\renewcommand{\arraystretch}{1.2}
\begin{tabular}{@{}|c|c|}
 \hline 
Process & Experimental bound \\ \hline \hline
$\mathrm{Br}\left[ \tau \to \mu \gamma \right] $ & ~$4.4 \times
10^{-8}$ ~\cite{Aubert:2009ag,Hayasaka:2007vc} \\
\hline 
$\mathrm{Br}\left[ \tau \to e \gamma \right] $ & ~$3.3 \times
10^{-8}$~ \cite{Aubert:2009ag} \\
\hline
$ \mathrm{Br}\left[ \mu \to e \gamma \right] $ & ~$5.7 \times
10^{-13}$~\cite{Adam:2013mnn} \\
\hline
$\mathrm{Br}\left[\tau^-\to\mu^-\mu^+\mu^-\right] $ & ~$2.1 \times
10^{-8}$ ~\cite{Hayasaka:2010np} \\
\hline 
$\mathrm{Br}\left[\tau^-\to e^-e^+e^-\right] $ & ~$2.7\times 10^{-8}$
~\cite{Hayasaka:2010np} \\
\hline 
$\mathrm{Br}\left[\tau^-\to e^- \mu^+\mu^- \right] $ & ~$2.7 \times
10^{-8}$~ \cite{Hayasaka:2010np} \\
\hline
$\mathrm{Br}\left[\tau^-\to \mu^- e^+\mu^- \right] $ & ~$1.7 \times
10^{-8}$~ \cite{Hayasaka:2010np} \\
\hline $ \mathrm{Br}\left[\mu^-\to e^-e^+e^- \right]$ & ~$1.0 \times
10^{-12}$~\cite{Bellgardt:1987du} \\
\hline $ \mathrm{Br}_{\rm Au}\left[\mu \to e \right]$ & ~$7.0 \times
10^{-13}$~\cite{Bertl:2006up} \\
\hline \hline
\end{tabular}
\end{minipage} 
\caption{Experimental upper limits on the branching ratios of LFV decays.}
\label{table:llllEXP}
\end{table} \textsc{}
\end{center}

Table \ref{table:llllEXP} shows the current experimental status of search for LFV decays.

LFV processes have been studied in great detail in many extensions of the SM. For example, in the MSSM non-vanishing rates for LFV processes are generated by flavor non-diagonal SUSY-breaking terms~\cite{Borzumati:1986qx}. Extending the MSSM with right-handed neutrinos by the seesaw mechanism gives rise to LFV~\cite{Ilakovac:1994kj}, as well as allowing for $R$-parity violation~\cite{deGouvea:2000cf}. The Littlest Higgs Model with $T$-parity~\cite{Blanke:2007db}, 2HDMs with generic flavor structures~\cite{Kanemura:2004cn}, and models with an extended fermion sector~\cite{Buras:2011ph} have sources of LFV as well.  In order to make NP scenarios consistent with the
non-observation of LFV processes in nature, the assumption of Minimal Flavor Violation has been extended to the lepton sector, see e.g.~\cite{Cirigliano:2005ck}. LFV decays have been studied in a model-independent way in~\cite{Raidal:1997hq}.

\begin{figure}
\begin{center}
\includegraphics[height=35ex]{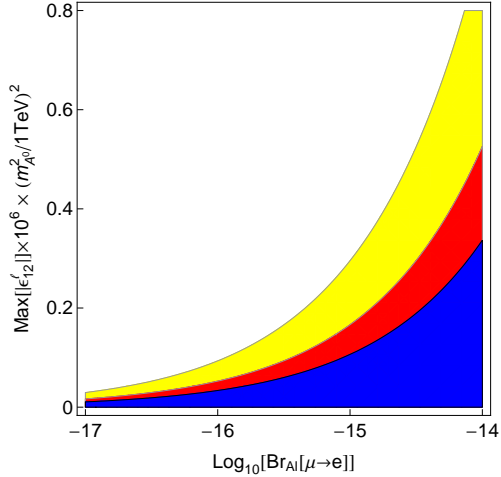}%
\end{center}
\caption{Allowed regions for $\epsilon^\ell_{12}\equiv\epsilon^\ell_{e\mu}$ as a function of the upper limit on $\mu\to e$ conversion in aluminum \cite{Crivellin:2014cta}. The blue, red, and yellow regions correspond to $\tan \beta = 50, 40, 30$, respectively (the regions are superimposed with more stringent limits for larger $\tan\beta$). Note the simple quadratic scaling of the constraints on the heavy Higgs mass. For the numerical evaluation we used the improved predictions of the nucleon-quark couplings of Ref.~\cite{Crivellin:2013ipa}.}
\label{fig:muecon}
\end{figure}

Let us consider $\mu\to e$ conversion in some more detail since it has very promising experimental prospect \cite{mueconversionEXP} and is especially suited to test Higgs mediated flavour violation since it does not necessarily involve couplings to light leptons as $\mu\to e \gamma$ \cite{Crivellin:2014cta}. In Fig.~\ref{fig:muecon} we show the constraints on $\epsilon^\ell_{12}$ in the 2HDM from $\mu\to e$ conversion.

\section{Deviations from the SM}
\label{sec:deviations}

\subsection{Tauonic $B$ decays}

Tauonic $B$-meson decays are an excellent probe of new physics: they test lepton flavor universality satisfied in the SM and are sensitive to new particles which couple proportionally to the mass of the involved particles (e.g. Higgs bosons) due to the heavy $\tau$ lepton involved. Recently, the BABAR collaboration performed an analysis of the semileptonic $B$ decays $B\to D\tau\nu$ and $B\to D^*\tau\nu$ using the full available data set \cite{BaBar:2012xj}. They find for the ratios
\begin{equation}
{\cal R}(D^{(*)})\,=\,{\cal B}(B\to D^{(*)} \tau \nu)/{\cal B}(B\to D^{(*)} \ell \nu)\,,
\end{equation}
the following results:
\begin{eqnarray}
{\cal R}(D)\,=\,0.440\pm0.058\pm0.042  \,,\qquad{\cal R}(D^*)\,=\,0.332\pm0.024\pm0.018\,.
\end{eqnarray}
Here the first error is statistical and the second one is systematic. Comparing these measurements to the SM predictions \cite{Korner:1987kd} (using the form factors of Refs~\cite{Dungel:2010uk,Amhis:2012bh,Beringer:1900zz})
\begin{eqnarray}
{\cal R}_{\rm SM}(D)\,=\,0.297\pm0.017 \,, \qquad{\cal R}_{\rm SM}(D^*) \,=\,0.252\pm0.003 \,,
\end{eqnarray}
we see that there is a discrepancy of 2.2\,$\sigma$ for $\cal{R}(D)$ and 2.7\,$\sigma$ for $\cal{R}(D^*)$ and combining them gives a $3.4\, \sigma$ deviation from the SM~\cite{BaBar:2012xj}. This evidence for new physics in $B$-meson decays to taus is further supported by the measurement of ${\cal B}[B\to \tau\nu]=(1.15\pm0.23)\times 10^{-4}$ which disagrees with by $1.6\, \sigma$ higher than the SM prediction using $V_{ub}$ from a global fit of the CKM matrix \cite{Charles:2004jd}.

The generic effect of NP including differential distributions has been studied in \cite{genericNPBDtaunu}. Many NP explanations of this anomaly have been proposed \cite{Fajfer:2012jt}.
A natural possibility to explain these enhancements compared to the SM prediction is a charged scalar particle which couples proportionally to the masses of the fermions involved in the interaction: a charged Higgs boson. A charged Higgs affects $B\to \tau\nu$~\cite{Hou:1992sy}, $B\to D\tau\nu$ and $B\to D^*\tau\nu$~\cite{Tanaka:1994ay}. In a 2HDM of type II (with MSSM like Higgs potential) the only free additional parameters are $\tan\beta=v_u/v_d$ (the ratio of the two vacuum expectation values) and the charged Higgs mass $m_{H^\pm}$ (the heavy CP even Higgs mass $m_{H^0}$ and the CP odd Higgs mass $m_{A^0}$ can be expressed in terms of the charged Higgs mass and differ only by electroweak corrections). In this setup the charged Higgs contribution to $B\to \tau\nu$ interferes necessarily destructively with the SM contribution\cite{Hou:1992sy}. Thus, an enhancement of $\cal B\left[B\to \tau\nu\right]$ is only possible if the absolute value of the charged Higgs contribution is bigger than two times the SM one. Furthermore, a 2HDM of type II cannot explain $\cal{R}(D)$ and $\cal{R}(D^*)$ simultaneously \cite{BaBar:2012xj}.

\begin{figure}[ht]
\centering
\includegraphics[width=0.3\textwidth]{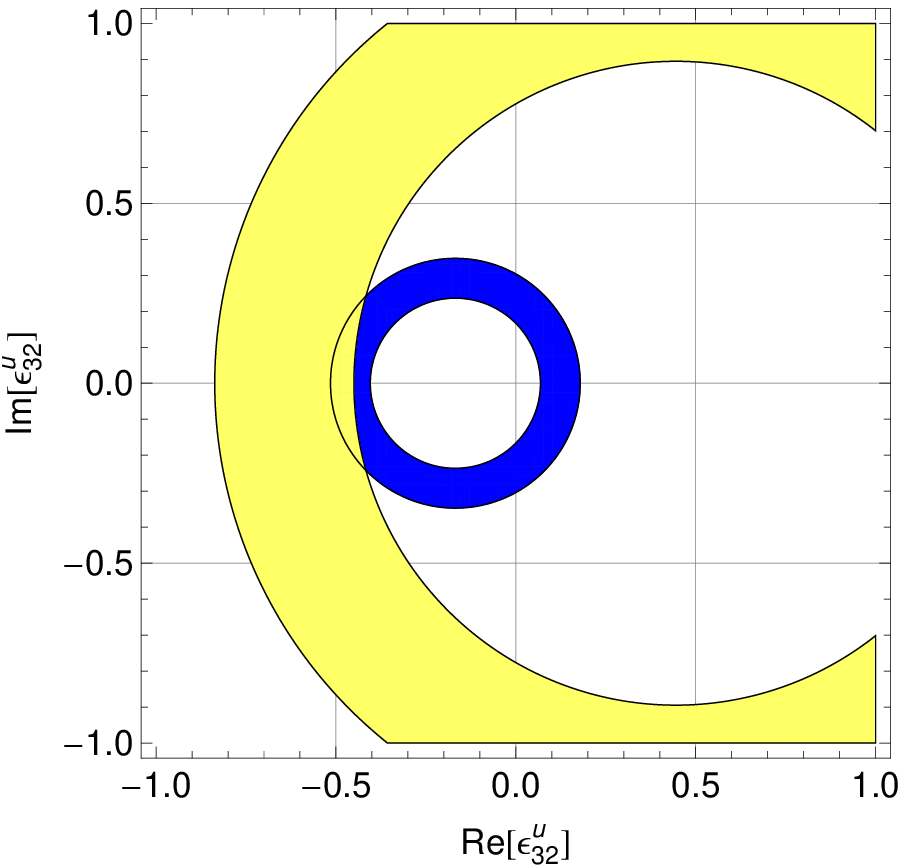}
\includegraphics[width=0.31\textwidth]{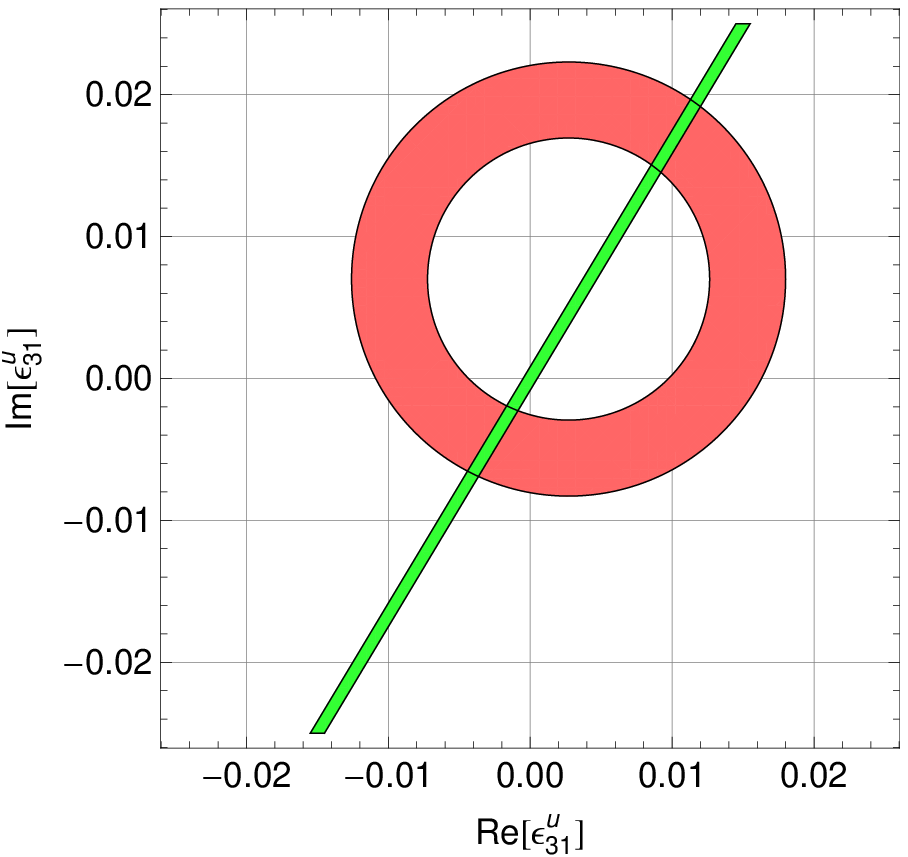}
\includegraphics[width=0.31\textwidth]{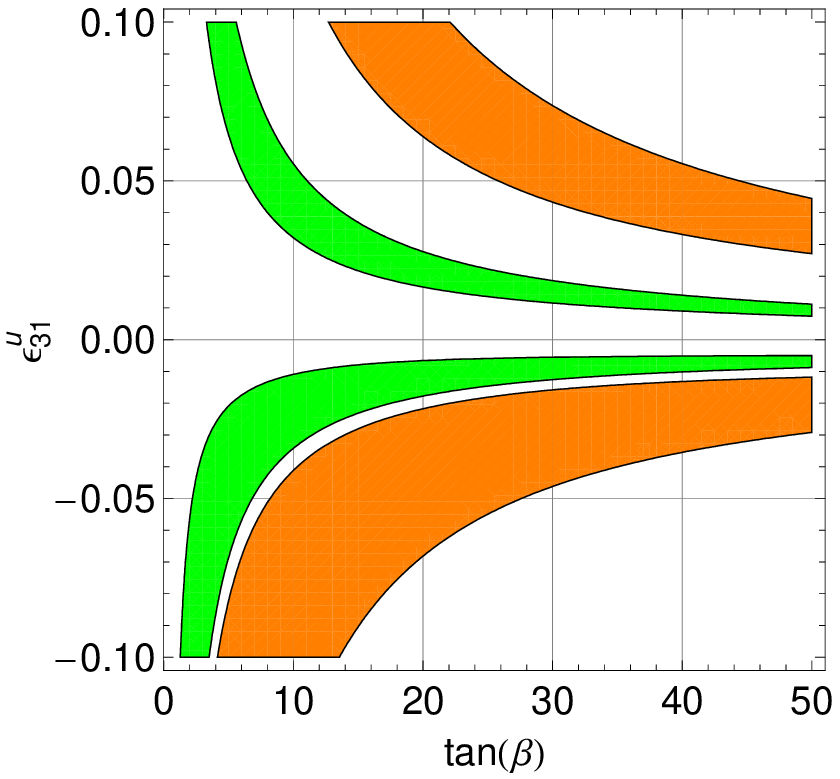}
\caption{Left: Allowed regions in the complex $\epsilon^u_{32}$-plane from $\cal{R}(D)$ (blue) and $\cal{R}(D^*)$ (yellow) for $\tan\beta=50$ and $m_H=500$~GeV. Middle:  Allowed regions in the complex $\epsilon^u_{31}$-plane from $B\to \tau\nu$. Right:  Allowed regions in the $\tan\beta$--$\epsilon^u_{31}$ plane from $B\to \tau\nu$ for real values of $\epsilon^u_{31}$ and $m_H=400$~GeV (green), $m_H=800$~GeV (orange). The scaling of the allowed region for $\epsilon^u_{32}$ with $\tan\beta$ and $m_H$ is the same as for $\epsilon^u_{31}$. $\epsilon^u_{32}$ and $\epsilon^u_{31}$ are given at the matching scale $m_H$. \label{2HDMIII}}
\end{figure}

As we found before, all $\epsilon^d_{ij}$ and $\epsilon^u_{13,23}$ are stringently constrained from FCNC processes in the down sector and only $\epsilon^u_{31}$ ($\epsilon^u_{32}$) significantly effects $B\to \tau\nu$ ($\cal{R}(D)$ and $\cal{R}(D^*)$) without any suppression by small CKM elements. Furthermore, since flavor-changing $t\to u$ (or $t\to c$) transitions are not constrained with sufficient accuracy, we can only constrain these elements from charged Higgs-induced FCNCs in the down sector. However, since in this case an up (charm) quark always propagates inside the loop, the contribution is suppressed by the small Yukawa couplings of the up-down-Higgs (charm-strange-Higgs) vertex involved in the corresponding diagrams. Thus, the constraints from FCNC processes are weak, and $\epsilon^u_{32,31}$ can be sizable. 
Indeed, it turns out that by using $\epsilon^u_{32,31}$ we can explain $\cal{R}(D^*)$ and $\cal{R}(D)$ simultaneously \cite{Crivellin:2012ye}. In Fig.~\ref{2HDMIII} we see the allowed region in the complex $\epsilon^u_{32}$-plane, which gives the correct values for $\cal{R}(D)$ and $\cal{R}(D^*)$ within the $1\, \sigma$ uncertainties for $\tan\beta=50$ and $M_H=500$~GeV. Similarly, $B\to \tau\nu$ can be explained by using $\epsilon^u_{31}$.

\boldmath
\subsection{$B\to K^*\mu^+\mu^-$ and $B\to K\mu^+\mu^-$ vs $B\to K e^+ e^-$}
\unboldmath

The decays $B\to K^*\ell^+\ell^-$ and $B\to K\ell^+\ell^-$ (with $\ell=e,\mu$) have been studied extensively in the SM (including also non-standard operator structures) \cite{BKmumuSMandbeyond}. 

While the forward-backward asymmetry in $B\to K^*\mu^+\mu^-$ agrees with the SM predictions \cite{Aaij:2013iag}, deviations from the SM predictions have been observed by LHCb in angular observables \cite{Aaij:2013qta}, mainly in the observable called $P_5^\prime$. While it is still possible that this anomaly could originate from hadronic uncertainties or power corrections \cite{BKmumuHadroinc}, it is still interesting to examine if and how NP can explain this anomaly. In an model independent approach, the deviations from the SM can be explained by rather large contributions to the Wilson coefficient $C_9$ \cite{Descotes-Genon:2013wba,Altmannshofer:2013foa}. Concerning concrete models of NP properbly the most natural expectation is a flavour changing $Z$ or $Z^\prime$ coupling \cite{Zprime} while explaining the central value of the anomaly in the MSSM is not possible without violating bounds from other observables \cite{Altmannshofer:2013foa,Mahmoudi:2014mja} which is also true for models with extra dimensions \cite{Biancofiore:2014wpa}.

Very recently, LHCb measured the ratio $R_K={\rm Br}[B\to K\mu^+\mu^-]/{\rm Br}[B\to K e^+ e^-]$ \cite{Aaij:2013qta} and found significant deviations from the SM prediction $R_K^{\rm SM}=1.0003 \pm 0.0001$:
\begin{equation}
R^{\rm LHCb}_K = 0.745^{+0.090}_{-0.074}\pm 0.036\,.
\end{equation}
A possible explanations would be NP contributing to $B\to K\mu^+\mu^-$ but not to $B\to K e^+e^-$ involving $C_9$ with muons only \cite{Hiller:2014yaa} which also give welcome NP effects in $B\to K^*\mu^+\mu^-$.

\section{Conclusions}

\begin{figure}[ht]
\begin{center}
\includegraphics[width=0.49\textwidth]{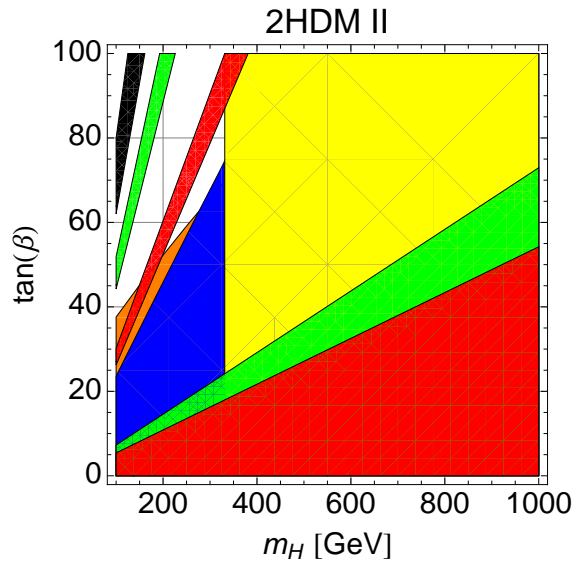}
\end{center}
\caption{Updated constraints on the 2HDM of type II parameter space \cite{Crivellin:2013wna}. The regions compatible with experiment are shown (the regions are superimposed on each other): $b\to s\gamma$ (yellow), $B\to D\tau\nu$ (green), $B\to \tau \nu$ (red), $B_{s}\to \mu^{+} \mu^{-}$ (orange), $K\to \mu \nu/\pi\to \mu \nu$ (blue) and $B\to D^*\tau \nu$ (black). Note that no region in parameter space is compatible with all processes. Explaining $B\to D^*\tau \nu$ would require very small Higgs masses and large values of $\tan\beta$ which is not compatible with the other observables. To obtain this plot, we added the theoretical uncertainty linearly on top of the $2 \, \sigma$ experimental error.}
\label{fig:2HDMII}
\end{figure}

In these proceedings I presented a personal perspective of the challenges for NP physics in the flavour sector. We reviewed the stringent constraints on physics beyond the SM imposed by flavour observables with must be respected by any viable model of NP. As an example, the combined flavour constraints on the 2HDM of type II are shown in Fig.~\ref{fig:2HDMII}. In case deviations from the SM are observed, it is interesting to examine which model of NP is capable of explaining such deviations without violating bounds from other observables. As examples we considered tauonic $B$ decays, $B\to K^*\ell^+\ell^-$ and $R_K$. While a rather natural explanation for $B\to (D^{(*)}\tau\nu)$ is a charged Higgs contribution in a 2HDM with flavour-changing couplings involving the top quark, $B\to K^*\ell^+\ell^-$ and $R_K$ can most naturally be explained by a $Z^\prime$ boson.

\Acknowledgements

I thank the organizers for the invitation and the possibility to present these results. I also thank Ulrich Nierste for useful discussions and proofreading. This work is supported by a Marie Curie Intra-European Fellowship of the European Community's 7th Framework Programme under contract number (PIEF-GA-2012-326948).

\end{document}